\documentclass[11pt]{article}
\usepackage{epstopdf}                                                             
\usepackage[a4paper,hmarginratio=1:1,vmarginratio=2:3,totalwidth=15.3cm,totalheight=22.cm]{geometry}
\usepackage{bm,epstopdf,epsfig,amsmath,amssymb,amsfonts,colordvi,latexsym,
  wrapfig,comment,cancel,verbatim,slashed}
\usepackage{epsfig,bbm}
\usepackage{graphicx,graphics}
\usepackage[font=md,captionskip=8pt]{subfig}
\usepackage[usenames,dvipsnames]{color}
\usepackage[noadjust]{cite}
\usepackage{xcolor} 
\usepackage[utf8]{inputenc}
\usepackage{setspace}

\usepackage[utf8]{inputenc}
\newcommand{\q}{\alpha}

\newcommand{\sg}{\sqrt{g}}    
\newcommand{\sgh}{\sqrt{\hat g}}
\newcommand{\w}{\omega}

\newcommand{\tGamma}{\tilde\Gamma}

\newcommand{\eff}{\textrm{eff}}
\allowdisplaybreaks

\newcommand{\cL}{{\cal L}}

\newcommand{\ra}{\rightarrow}

\newcommand{\be}{\begin{equation}}
\newcommand{\ee}{\end{equation}}
\newcommand{\bea}{\begin{eqnarray}}
\newcommand{\eea}{\end{eqnarray}}
\newcommand{\Ra}{\Rightarrow}

\newcommand{\baa}{\begin{array}}
\newcommand{\eaa}{\end{array}}

\long\def\symbolfootnote[#1]#2{\begingroup
\def\thefootnote{\fnsymbol{footnote}}\footnote[#1]{#2}\endgroup}

\begin{document}
\begin{flushright}
\end{flushright}
\bigskip\medskip
\thispagestyle{empty}
\vspace{1cm}

\begin{center}
\vspace{1cm}

 {\Large \bf  Cosmological evolution in  Weyl conformal geometry} 

 \vspace{1.5cm}
 
 {\bf D. M. Ghilencea}$^{\,\,a}$\,\symbolfootnote[1]{E-mail: dumitru.ghilencea@cern.ch, t.harko@ucl.ac.uk}
  and {\bf T. Harko}$^{\,\,a,b,c}$
\bigskip

$^a$ {\small Department of Theoretical Physics, National Institute of Physics
 \smallskip 

 and  Nuclear Engineering (IFIN), Bucharest, 077125 Romania}

\smallskip

$^b$ {\small Department of Physics, Babe\c{s}-Bolyai University, Cluj-Napoca, 400084 Romania}

\smallskip

$^c$ {\small Astronomical Observatory, Cluj-Napoca, 400487 Romania}

\end{center}

\medskip

\begin{abstract}
  \begin{spacing}{1.}
\noindent
    We discuss the cosmological evolution of the  Weyl conformal geometry and its
    associated Weyl quadratic gravity.  The Einstein gravity
    (with a positive cosmological constant)  is recovered in  the spontaneously
    broken phase of Weyl  gravity; this happens after  the  Weyl gauge field
 ($\w_\mu$) of scale symmetry, that is part of the Weyl geometry,
 becomes massive by Stueckelberg mechanism and decouples. This breaking is a natural result
 of the cosmological evolution of Weyl geometry, in the absence of matter.
 Of particular interest in the analysis is the special limiting case of Weyl integrable geometry.
 Both this case as well as  the general one provide
 an accelerated expansion of the Universe, controlled  by the scalar mode of the
 $\tilde R^2$ term in the action and by $\w_0$.
Their  comparison to  the  $\Lambda$CDM model  shows a very good agreement to this model
for the (dimensionless) Hubble function $h(z)$ and the  deceleration $q(z)$
 for redshift $z\leq 3$. Therefore, the Weyl conformal geometry and its associated
 Weyl quadratic gravity  provide  an  interesting  alternative to the $\Lambda$CDM model
 and to the Einstein gravity. 
\end{spacing}
\end{abstract}

\newpage

 \section{Introduction}

 The original Weyl quadratic gravity based on the Weyl conformal geometry \cite{Weyl1,Weyl2,Weyl3}
 provided an alternative to Einstein's general relativity.
 The theory has a {\it gauged} scale symmetry known as  Weyl gauge symmetry that follows from the
 underlying Weyl geometry. Hence, both the action {\it and} the geometry (connection)
 have this  symmetry. The Weyl quadratic gravity was soon disregarded by  Einstein's
 critique of its non-metricity \cite{Weyl1}\footnote{We ignore here
   Weyl's unfortunate wrong interpretation of $\w_\mu$ as the real photon.}
 related to the  (apparently massless) dynamical  Weyl gauge field $\w_\mu$
\footnote{Actually, quadratic gravity in  so-called ``Palatini approach''  due to Einstein \cite{E1,E2}
  is also non-metric \cite{Ghilen2,Winflation2}.}. Dirac   revived the interest
in this theory  but considered Weyl's quadratic action ``too complicated to be satisfactory''
and introduced instead  a  simplified, linear  version of this theory \cite{Dirac}.
Subsequent studies followed this approach \cite{Kugo,
   Smolin,Cheng,Fulton,Aluri, Moffat1,Nishino,Ohanian,Moffat2,Tann,Huang,
   ghilen,Guendelman,pp1,pp2,pp3,pp4,pp5} (for a review  see \cite{Scholz}).
 These models are limited to Lagrangians {\it linear} in the
scalar curvature $\tilde R$ of Weyl geometry and thus need
{\it additional states} (scalar fields beyond the Higgs)  to maintain the Weyl
gauge symmetry   and to generate the  mass scales of the theory (Planck,  etc)
by vacuum expectation values (vev's) of these scalar fields.

The {\it original} Weyl quadratic gravity was re-considered in \cite{Ghilen1,SMW}  where it was shown 
that the  Weyl  field  $\w_\mu$ is actually {\it massive}, possibly near the Planck scale ($M_p$),
after a geometric Stueckelberg mechanism; in  this,  $\w_\mu$ ``absorbs'' the
scalar field ($\phi_0$) extracted from the $\tilde R^2$ term in the action.
Hence,  non-metricity is not a problem  since it is suppressed by the
(large) mass of the Weyl gauge boson  and the theory is then viable.
 Actually, it is the non-metricity of the underlying geometry that ensures a spontaneous
 breaking of  the Weyl gauge symmetry and the  mass  generation,  like Planck scale or
 the Weyl  field mass ($m_\w$). This breaking takes place in the absence of matter fields.
 Below $m_\w$, the field $\w_\mu$ decouples, to restore metricity
 and leave {\it in the broken phase} the Einstein gravity and a positive cosmological constant.
 For ultraweak  coupling, $m_\w$ may be much lighter, of
 few TeV, which is the current lower bound on the non-metricity scale
 \cite{Latorre}. Similar results exist in  Palatini quadratic gravity \cite{Ghilen2,Winflation2}
 and may apply to  metric affine gravity   \cite{Percacci,P1,P2}.

 The Weyl gauge symmetry  is  preferable since it has a non-trivial current
 \cite{Ghilen1,SMW} which is unlike the  Weyl  local symmetry (without $\w_\mu$) \cite{J1,J2}),
and it  has  a geometric interpretation (in Weyl geometry)
which does not seem possible for Weyl symmetry (without $\w_\mu$) \cite{Ohanian,Quiros1,Quiros2}.
It is also preferable to the {\it global} scale symmetry which is broken by black-hole
physics \cite{RK}. The (geometric)  field $\w_\mu$ may also bring  a geometric solution
to the dark matter problem.

  Interestingly, Weyl geometry provides a natural embedding of the
  Standard Model (with a vanishing Higgs mass)   {\it without}   new degrees of freedom required 
  beyond the SM spectrum and Weyl geometry \cite{SMW}.
  Mass generation (Higgs vev, $M_p$, fermions masses, etc)
  then follow from  the  Stueckelberg  breaking  of the Weyl gauge symmetry. 
 Models in Weyl geometry also have successful inflation  \cite{Winflation,Ross,Winflation2} with
 predictions similar to  those in the  Starobinsky model \cite{Star}.
 Briefly,  Weyl geometry is a viable frame  for model building beyond the
 SM that {\it  automatically  includes the Einstein gravity} and
 a positive cosmological constant.

 Motivated by these results, here we consider the cosmological evolution
  in Weyl  geometry and its associated  Weyl quadratic gravity.
  The study continues that  in  \cite{Ghilen1,SMW} at the level of the equations of motion
  and   sheds new light  on the spontaneous breaking of the Weyl gauge symmetry:
  we show (Section~\ref{s2}) that the breaking of the symmetry and the
  ``gauge fixing'' condition ($\nabla_\mu\w^\mu\!=\!0$) specific to a massive
  gauge field,  are a  natural result of the cosmological evolution in a
  Friedmann-Lema\^itre-Robertson-Walker (FLRW) universe.
  An interesting aspect is that  all mass scales of the theory
  (Planck scale $M_p$, cosmological constant $\Lambda$,  $m_\w$) have a geometric origin,
  due to $\phi_0$ propagated by the $\tilde R^2$ term. We  then show
  (Section~\ref{s3})  that, in the absence of matter, the Weyl geometry and
  its associated Weyl quadratic gravity provide  an accelerated expansion of
  the Universe in agreement with recent results \cite{1n,2n,3n,4n,acc,Aad,Rie}.
  The scalar mode (Stueckelberg field $\phi_0$) in the $\tilde R^2$ term of the action
  contributes (positively)
 to this acceleration together with the time-like component of $\w_\mu$ which also gives 
 a dark matter-like contribution (of opposite sign).

 A particularly interesting case is the limit of a Weyl integrable geometry when
 $\w_\mu$ is ``pure gauge'', giving an isotropic solution.
 This case is discussed and compared numerically to the general case.
 A good agreement is found  of both these Weyl cases  with
 the  $\Lambda$CDM model (based on the Einstein gravity with a cosmological
 constant)   for the Hubble function  $h(z)$ and the deceleration $q(z)$
 (Section~\ref{s3}). These results indicate that the Weyl conformal geometry and its
 associated  Weyl quadratic gravity can provide an interesting alternative
 to the $\Lambda$CDM model and to the Einstein gravity.
 This  suggests that, ultimately,
 the underlying geometry of our Universe may actually  be the Weyl conformal geometry.
 Our conclusions are  found in Section~\ref{s4}.

\section{Weyl action and spontaneous  symmetry breaking}\label{s2}

\subsection{Brief review of Weyl action}

 Weyl geometry is defined by classes of equivalence
$(g_{\alpha\beta}, \w_\mu$) of the metric ($g_{\alpha\beta}$)
and the Weyl gauge  field ($\w_\mu$),  related by the  Weyl gauge
transformation, see $(a)$ below. If matter is present, $(a)$ must be extended by
transformation  $(b)$ of the scalars ($\phi$) and fermions ($\psi$)
\bea\label{WGS}
 (a) &\quad&
 \hat g_{\mu\nu}=\Sigma^d
 \,g_{\mu\nu},\qquad
\hat\w_\mu=\w_\mu -\frac{1}{\q}\, \partial_\mu\ln\Sigma, 
\qquad
\sqrt{\hat g}=\Sigma^{2 d} \sqrt{g},
\nonumber\\[3pt]
(b) &\quad & \hat \phi = \Sigma^{-d/2} \phi, \qquad \hat\psi=\Sigma^{-3d/4}\,\psi,
\qquad\qquad\quad (d=1).
\eea
Here $d$ is the Weyl charge of $g_{\mu\nu}$,
$\alpha$ is the Weyl gauge coupling\footnote{Our convention
  is  $g_{\mu\nu}=(+,-,-,-)$   while the curvature tensors are defined as in  \cite{book}.},
$g\!=\!\vert\det g_{\mu\nu}\vert$; $\Sigma\!>\!0$; without loss of generality, we set $d\!=\!1$.
The Weyl connection $\tGamma$ is a solution to
$\tilde\nabla_\lambda  g_{\mu\nu}=-  \q\, \w_\lambda g_{\mu\nu}$
where $\tilde\nabla_\mu$ is defined by  $\tilde\Gamma_{\mu\nu}^\lambda$, with:
\be\label{tGamma}
\tilde \Gamma_{\mu\nu}^\lambda=
\Gamma_{\mu\nu}^\lambda+(1/2)\,\q \,\Big[\delta_\mu^\lambda\,\, \w_\nu +\delta_\nu^\lambda\,\, \w_\mu
- g_{\mu\nu} \,\w^\lambda\Big], \qquad \Ra\quad
\w_\mu \propto \tilde\Gamma_\mu-\Gamma_\mu,
\ee
with the notation $\Gamma_\mu=\Gamma_{\mu\nu}^\nu$, $\tilde\Gamma_\mu=\tilde\Gamma_{\mu\nu}^\nu$.
Hence, the Weyl gauge field  $\w_\mu$ measures the departure of the (trace of) the Weyl
connection $\tilde \Gamma$ from the Levi-Civita connection $\Gamma$. If $\w_\mu$ is massive
and decouples ($\w_\mu\ra 0$),  $\tilde \Gamma\ra \Gamma$ and Weyl geometry becomes Riemannian.

With $\tilde\Gamma$ of
(\ref{tGamma}) one defines the tensor and scalar curvature of Weyl geometry, via
the usual formulae of the Riemannian case, see e.g. Appendix A in \cite{SMW}.
With this, one finds that the scalar curvatures $\tilde R$ of  Weyl geometry
and $R$ of the Riemannian  geometry are related by
\be
\tilde R=R - 3\, \alpha\, \nabla_\mu\w^\mu - \frac32 \,\alpha^2\, \w_\mu \w^\mu.
\label{def}
\ee
The rhs of (\ref{def}) is in a Riemannian notation, so
$\nabla_\mu\w^\lambda=\partial_\mu \w^\lambda+\Gamma^\lambda_{\mu\rho}\,\w^\rho$.
The advantage of Weyl geometry is that
$\tilde R$ transforms covariantly, just like the square
of a scalar field\footnote{$\cL_0$ may also contain an additional term due to the
Weyl tensor of Weyl geometry ($\tilde C_{\mu\nu\rho\sigma}^2$)
\vspace{-0.1cm}\be
\cL_0^\prime=- (\sqrt{g}/\eta^2) \tilde C_{\mu\nu\rho\sigma}^2,\qquad \eta<1,
\quad\textrm{where}\quad
\tilde C_{\mu\nu\rho\sigma}^2= C_{\mu\nu\rho\sigma}^2+ (3/2) \,\q^2\,F_{\mu\nu}^2,
\ee

\vspace{-0.1cm}\noindent
where $\tilde C_{\mu\nu\rho\sigma}$  is related to its
Riemannian counterpart $C_{\mu\nu\rho\sigma}$ as shown above.
This term may be needed at a quantum level and brings a ghost degree of freedom
and a renormalization of the coupling of the  $F^2$ term.  We do not include this term in our present
study. For an analysis of the  $C_{\mu\nu\rho\sigma}^2$  term see \cite{Ma0,Ma1,Ma2,Ma3,Ma4,Ma5,Ma6,Ma7}.
}.

The gravity action in Weyl geometry was introduced in
\cite{Weyl1,Weyl2,Weyl3} and here we follow \cite{Ghilen1}
\bea\label{inA}
\cL_0=\sg\, \,\Big[\, \frac{1}{4!}\,\frac{1}{\xi^2}\,\tilde R^2  - \frac14\, F_{\mu\nu}^{\,2} \Big],
\eea
with perturbative coupling  $\xi < 1$. Here
$F_{\mu\nu}=\tilde\nabla_\mu\w_\nu-\tilde\nabla_\mu\w_\nu$ is the field strength of $\w_\mu$,
with $\tilde\nabla_\mu\w_\nu=\partial_\mu\w_\nu-\tilde\Gamma_{\mu\nu}^\rho\w_\rho$.
Since  $\tilde\Gamma_{\mu\nu}^\alpha\!=\!\tilde\Gamma_{\nu\mu}^\alpha$ is symmetric,
then $F_{\mu\nu}=\partial_\mu\w_\nu-\partial_\nu\w_\mu$.

To simplify the calculations,
in $\cL_0$ one can replace  $\tilde R^2\ra -2 \phi_0^2\,\tilde R-\phi_0^4$
where $\phi_0$ is a scalar field. This gives a {\it classically  equivalent} Lagrangian since
by using the solution $\phi_0^2=-\tilde R$ of the equation of motion of $\phi_0$ back in the modified
$\cL_0$, one recovers onshell eq.(\ref{inA}). Hence
\bea\label{alt3}
\cL_0=\sqrt{g}\, \Big[-\frac{1}{12}\frac{1}{\xi^2}\,\phi_0^2\,\tilde R
-\frac{\phi_0^4}{4!\,\xi^2}
-\frac14 \,F_{\mu\nu}^2
\Big].
\eea

\medskip\noindent
This is the simplest action with Weyl gauge symmetry that we shall use, equivalent to (\ref{inA}).
The advantage of (\ref{alt3}) over (\ref{inA})  is that the equations of motion simplify considerably
since  (\ref{alt3}) is  now {\it linear} in the curvature while  the  field $\phi_0$ becomes
dynamical, see later\footnote{Similar to the Riemannian case,
  $\tilde R^2$ propagates a  spin-zero mode ($\phi_0$) beyond the graviton, 
  because  it contains the higher derivative $R^2$, see (\ref{tGamma}), (\ref{def}); 
  that $\phi_0$ is dynamical is also seen from its eq of motion, eq.(\ref{box}).}.

\subsection{From Weyl to Einstein}\label{WtoE}

As shown in \cite{Ghilen1,SMW}, $\cL_0$  has  spontaneous
breaking to an Einstein-Proca Lagrangian of the Weyl gauge field. 
Here we briefly review this result.
In $\cL_0$ replace  $\tilde R$ by  eq.(\ref{def}).
\bea\label{alt}
\cL_0=\sqrt{g}\,\Big\{- \frac{1}{12}\,\frac{\phi_0^2}{\xi^2}
\,\Big[ R- 3 \q \nabla_\mu\w^\mu -\frac{3}{2}
\q^2\,\w_\mu\w^\mu\Big]
-\frac{1}{4!}\frac{\phi_0^4}{\xi^2} - \frac14 \,F_{\mu\nu}^2
\Big\}.
\eea

\smallskip\noindent
This can be re-written as
\smallskip
\be\label{alt2}
\!\cL_0\!=\!\sqrt g\,
\Big\{\frac{-1}{2\,\xi^2}\,
\Big[ \frac{\phi_0^2}{6}\, R\,
+(\partial_\mu\phi_0)^2
-\,\frac{\q}{2}\,\nabla_\mu (\w^\mu\phi_0^2)
\Big]
-\frac{\phi_0^4}{4!\,\xi^2}
+\frac{\q^2}{8\,\xi^2}\,\phi_0^2\, \Big[\w_\mu -\frac{1}{\q}\partial_\mu \ln\phi_0^2\Big]^2\!
-\frac{1}{4}\, F_{\mu\nu}^2\Big\}
\ee

\smallskip\noindent
Each term multiplied by $1/\xi^2$ and  $\cL_0$
are invariant under  (\ref{WGS}). One would like to  ``fix the gauge'' of this Weyl gauge symmetry.
To do so,  apply to $\cL_0$   transformation (\ref{WGS}) with a scale-dependent
$\Sigma=\phi_0^2/\langle\phi_0^2\rangle$; this is  fixing $\phi_0$ to its vev;
naively, one simply sets $\phi_0\ra \langle\phi_0\rangle$ in eq.(\ref{alt2}).
We discuss shortly (Section~\ref{2.3}) how $\phi_0$ acquires a vev and how this gauge is fixed
by the cosmological evolution.
In terms of the new,  transformed fields (with a ``hat''),  $\cL_0$ becomes
\be
\label{EP}
\cL_0=\sgh \,\Big[- \frac12 M_p^2\hat R +\frac34 M_p^2\,\q^2\,\hat\w_\mu \hat \w^\mu
-\Lambda\,M_p^2
-\frac{1}{4} \, \hat F_{\mu\nu}^2 \Big],
\quad
M_p^2\equiv \frac{\langle\phi_0^2\rangle}{6\,\xi^2};\,\,\, \Lambda\equiv\frac14 \langle\phi_0^2\rangle.
\ee

\smallskip\noindent
where we ignored a total derivative  in the action.
This is the Einstein-Proca Lagrangian for the Weyl vector \cite{Ghilen1,SMW},
in the Einstein gauge ("frame").  The Weyl gauge field
has absorbed the derivative of the field $\ln\phi_0$ in a Stueckelberg mechanism:
the massless $\w_\mu$ and real, massless $\phi_0$ are replaced by a massive Weyl
gauge field, with a mass $m_\w^2=(3/2) M_p^2 \q^2$.
This mass is close to the Planck scale, unless
one is tuning $\q\ll 1$; hence, any non-metricity effects are strongly suppressed by $m_\w\sim M_p$.
Current lower bounds on this mass (which sets the non-metricity scale)
are actually very mild, close to the TeV scale
\cite{Latorre}. Since $\w_\mu$ is massive it can now  decouple in eq.(\ref{EP})
to leave in the broken phase  
(below $m_\w$) the Einstein action with a positive cosmological constant.
Hence,  the Einstein action is a {\it ``low energy'' broken phase  limit of the original Weyl quadratic
gravity}.  At the same time the connection $\tGamma$ of (\ref{tGamma}) becomes
Levi-Civita ($\Gamma$) and the geometry becomes Riemannian.
All mass scales of the theory ($M_p$, $m_\w$,  $\Lambda$) have {\it geometric origin}, being
proportional to  the vev of $\phi_0$ propagated by the $\tilde R^2$ term in the action.
For  details  see  \cite{Ghilen1,SMW}.

An interesting limit of Weyl geometry is the case  $\w_\nu= (1/\alpha)\  \partial_\nu\ln\phi_0^2$
i.e. $\w_\mu$ is actually ``pure gauge''.
This is the so-called Weyl integrable limit of the  Weyl geometry action considered\footnote{
  The cosmological implications of the Weyl integrable geometry were considered 
  in \cite{Bellini}, while for the analysis of other physical and geometrical
  aspects of the theory see \cite{Barrow1, Barrow2}.}. 
In this case the kinetic term of $\w_\mu$ is  vanishing, so the action is then given by the first term in
eq.(\ref{inA}).
Then from eq.(\ref{alt3}) without the last term,  one can analytically integrate out $\w_\mu$ 
and finds
\bea\label{WIG}
\cL_0=\sqrt{g} \frac{1}{\,\xi^2}
\Big\{
\frac{-1}{2}\Big[
\frac16\,\phi^2_0 R   + g^{\mu\nu}\,\partial_\mu\phi_0\partial_\nu\phi_0\Big] -\frac{1}{4!}
\,\phi_0^4\Big\}.
\eea
%
The action has now a  Weyl local symmetry only (no $\w_\mu$) (see also (\ref{WGS}) with $\hat\w_\mu=0$
for suitable $\Sigma$). Notice that, just like in the general case, 
after gauge fixing\footnote{We discuss in the next section how $\phi_0$ acquires a non-zero
  vev.} in which $\phi_0\ra \langle\phi_0\rangle$, one obtains
the usual Einstein term and also the cosmological constant term from last term in (\ref{WIG}),
with a positive sign.
The cosmological constant comes from the scalar mode ``extracted'' from $\tilde R^2$, just
like in the general case. So both the Einstein action
and a  positive cosmological constant are obtained in the broken phase
and originate from the initial\footnote{A similar
  conclusion applies to $R^2$ gravity in the Palatini formalism, see  \cite{Ghilen2} (Section 2) and
  \cite{NMGM}.} $\tilde R^2$.
Hence this limiting case is not really conformal to the Einstein action (where
$\Lambda$ can be  added with arbitrary sign and size).

An action and results similar   to (\ref{WIG}) are also obtained
from the general case  when the mass of $\w_\mu$ is large enough, $\sim$ Planck scale;
then  $\w_\mu$ can be integrated out and one obtains again action (\ref{WIG}) after gauge fixing,
up to corrections suppressed by $M_p$.

\subsection{Equations of motion}\label{2.3}

The above breaking of Weyl gauge symmetry discussed at the level of the Lagrangian
can also be understood from the equations of motion,  as a natural result of
the cosmological evolution in an FLRW universe. To this purpose,
let us write the equations of motion for our action, eq.(\ref{alt}).
To simplify the notation below, let us denote:
\bea
K=\frac{\phi_0^2}{\xi^2}, \qquad V=\frac{1}{4!} \,\frac{\phi_0^4}{\xi^2}.
\eea
Variation of (\ref{alt})  with respect to the metric gives
\bea\label{eqg}
\frac{1}{\sqrt{g}}
\frac{\delta \cL_0}{\delta g^{\mu\nu}}
&=&-\frac{1}{12} \,K\,\Big( R_{\mu\nu}-\frac12 \,g_{\mu\nu} \,R\Big)
+\frac{1}{12}\,\Big(  g_{\mu\nu}\Box-\nabla_\mu \nabla_\nu\Big)  K
\nonumber\\
&-&\frac{\q^2}{16}\, K \Big( g_{\mu\nu}\,\w^\rho\,\w_\rho - 2 \w_\mu\,\w_\nu\Big)
+\frac{\q}{8}\, K\, \Big( \nabla_\mu\w_\nu+\nabla_\nu \w_\mu- g_{\mu\nu}\,\nabla_\rho\w^\rho\Big)
\nonumber\\[4pt]
&+&\frac12 \,g_{\mu\nu}\,V
+\frac12\,\Big(\frac14 \, g_{\mu\nu}\,F_{\alpha\beta}\,F^{\alpha\beta} -g^{\alpha\beta}\,F_{\mu\alpha}
F_{\nu\beta}\Big).
\eea
Taking the trace gives
\bea\label{eqt}
\frac{1}{12}\, K\,R+\frac14 \,\Box K -\frac{\q^2}{8}\,K\,\w_\rho\,\w^\rho
-\frac{\q}{4} K\,\nabla_\rho  \,\w^\rho+2 V=0.
\eea
%
The equation of motion of $\phi_0$ is
\bea\label{eqf}
\frac{1}{12}\, K\, R -\frac{\q^2}{8} \,K\,\w_\rho\,\w^\rho-\frac{\q}{4}\,K\,\nabla_\rho\,\w^\rho +
\frac12\,\phi_0 \,\frac{\partial V}{\partial \phi_0}=0.
\eea
On the ground state this gives
$\langle\phi_0^2\rangle=-\tilde R=- [ R- (3/2) \q^2 \w_\rho \,w^\rho]$, which we already know.

The equation of motion of $\w_\mu$ is
\bea\label{eqw}
\frac{\q^2}{4 } K\,\w^\rho -\frac{\q}{4} \,g^{\rho\sigma}
\,\nabla_\sigma K +\nabla_\sigma F^{\rho\sigma}=0.
\qquad\qquad\quad
\eea
From eqs.(\ref{eqt}), (\ref{eqf}) then
\bea\label{box}
\Box K=0,\qquad\Ra\qquad
\partial^\mu(\sqrt{g}\,\, \partial_\mu \phi_0^2)=0.
\eea

\medskip\noindent
where $\Box=\nabla^\mu\nabla_\mu$.
There is thus an onshell conserved current $K_\mu\equiv \sqrt{g}\,\, \partial_\mu\phi_0^2$.
The equation of motion of  $\phi_0$, eq.(\ref{box}),  is  non-trivial
showing that  this field, after  ``linearising''  $\tilde R^2$, became
dynamical and corresponds to the spin-zero mode that $\tilde R^2$
propagates beyond the graviton (similar to the  Riemannian $R^2$ that it actually contains).

From the equation of motion of $\w_\mu$ by applying $\nabla_\sigma$ 
we  find a conserved current\footnote{
In the global scale invariant case there is a  non-trivial current
 \cite{Fe0,Fe1,Fe2,Fe3}, as above but with $\w_\mu=0$.}
%
\bea\label{JJ}
\nabla_\mu J^\mu=0,\qquad
J^\mu=-\frac{\q}{4} g^{\mu\nu}\big(\partial_\nu-\q\,\w_\nu\big) K
=-\frac{\q}{4\,\xi^2} g^{\mu\nu}\big(\partial_\nu-\q\,\w_\nu\big) \phi_0^2,
\eea
%
where we used the antisymmetry of $F_{\mu\nu}$. But using that $\Box K=0$ we also find
\bea\label{wc}
\nabla_\mu J^\mu=\frac{1}{\sqrt{g}} \partial_\mu (\sqrt{g}\,J^\mu)
=\frac{\q^2}{4 \xi^2}\,\sqrt{g} \,\nabla_\mu (\w^\mu \,\phi_0^2).
\eea
Hence $\nabla_\mu(\phi_0^2\w^\mu)=0$ which will
be used in Section~\ref{GFSB}.
Notice that in the Weyl integrable limit $\w_\mu=(1/\q) \partial_\mu\ln\phi_0^2$ the current
is vanishing, while $\nabla_\mu(\phi_0^2\w^\mu)=0$ becomes $\Box\phi_0^2=0$ which is
already seen in (\ref{box}).

\subsection{Gauge fixing and symmetry  breaking by cosmological evolution}
\label{GFSB}

Consider hereafter the
FLRW metric $g_{\mu\nu}=(1, - a(t)^2, -a(t)^2, -a(t)^2)$ and with $\phi_0=\phi_0(t)$ only,
then eq.(\ref{box})  can be written as below, with $H=\dot a/a$:
\bea\label{Keq}
{\ddot K}+ 3 H \dot K=0.
\eea
This gives
\bea\label{solp}\dot\phi_0=\frac{c_0}{a(t)^3\,\phi_0}, \quad \textrm{and}\quad
\phi^2_0(t)= c_0 \int_0^t \frac{d\tau}{a(\tau)^3} +c_1,
\eea
where $c_{0,1}$ are some constants. 
For $t\ra \infty $, similar to a global Weyl symmetry  \cite{Fe0,Fe1,Fe2,Fe3},
$\phi_0$ evolves to a constant. What happens then  to this degree of freedom? 
In this case we find from eq.(\ref{wc}) a gauge fixing condition
specific to massive gauge fields
\bea
\nabla_\mu \w^\mu=0.
\eea
This means that the field $\w_\mu$ has become a massive Proca field, by ``absorbing''  the $\phi_0$
degree of freedom which thus disappears from the spectrum.
To see this in more detail, consider the quasi-homogeneous case of 
$\partial_i \,\w_\mu=0$ \footnote{The other possibility consistent with a FLRW metric is
  $\w_i(t)=0$, with $i=1,2,3$, see next section.}.
Eq.(\ref{eqw})  for $\w_\mu$ gives for temporal $\mu=0$ and spatial $\mu=i$ components
\bea\label{ww}
\frac{\q}{2} \,\w_0=\partial_0 \ln\phi_0,
\qquad\qquad
\ddot \w_i+ H\,\dot \w_i+ \frac{\q^2}{4\,\xi^2}\,\w_i \,\phi_0^2=0,\qquad\quad
\eea
The solution for $\w_0(t)$ is
\bea
\w_0(t)=2 \,c_0\, \big[\,\q \,a(t)^3\,\phi_0(t)^2\,\big]^{-1}
\eea
When $\phi_0$ becomes a constant (vev),  $\w_0\ra 0$ while 
$\w_i$ satisfies eq.(\ref{ww})  but with $\phi_0\ra\langle\phi_0\rangle$:
\bea\label{ww2}
\ddot \w_i+ H\,\dot \w_i+ \frac{\q^2}{4\,\xi^2}\,\w_i \,\langle\phi_0^2\rangle=0.\qquad\quad
\eea
Hence $\w_i$ satisfies the equations  of on oscillator with a  mass\footnote{Since
  the equation of motion is linear in $\w_\mu$, the perturbations about $\w_\mu(t)$ respect
  the same relation.}
\bea
m_\w^2=\frac{\q^2}{4\,\xi^2}\,\langle\phi_0^2\rangle.
\eea
This result is  also obvious from eq.(\ref{eqw}) with  $\phi_0\ra \langle\phi_0\rangle$,
which shows a Proca field equation.
From (\ref{alt}) the Planck scale 
\be
M^2_p=\frac{\langle\phi_0\rangle^2}{6\,\xi^2},\quad \Ra\quad
m_\w^2=\frac{3\q^2}{2}\,M_p^2,
\ee
 in agreement with the mass of $\w_\mu$ shown in the Lagrangian of eq.(\ref{EP}).

 Therefore, the breaking of the symmetry,  Proca mass generation for $\w_\mu$  and the
 gauge fixing of the Weyl gauge symmetry are natural results of the cosmological
 evolution in the FLRW universe. After $\w_\mu$ decouples the connection (\ref{tGamma}) evolves
into  Levi-Civita and then the geometry becomes  Riemannian.
These results, obtained from the equations of motion,
complement the Lagrangian picture reviewed in Section~\ref{WtoE}.
This breaking mechanism is entirely geometrical:
 there is no scalar field added to this purpose: $\phi_0$ has geometrical
 origin, from the $\tilde R^2$ term, while $\w_\mu$ is an intrinsic part of the underlying
 Weyl geometry.

Further, the solution to (\ref{ww2}), assuming $H\approx$ constant,  is of the form
\bea\label{sol}
\w_i(t)=\frac{1}{\sqrt a}\big[ A \cos\theta(t) + B \sin\theta(t)\big],\qquad \theta(t)=\gamma \,t,
\quad \gamma^2=m_\w^2-\frac14 \,H^2.
\eea
with  $A, B$  constants; for $m_\w^2\gg H^2$  this solution oscillates rapidly.
In the general case ($H$ not constant)  $A, B$ become functions of time. This will
be used in cosmological applications.

Finally, note that  eq.(\ref{eqg}) also gives for $i\not=j$ that 
\medskip
\bea\label{ee}
\frac{\q^2}{4}\,K\,\w_i\,\w_j=\dot\w_i\,\dot\w_j,\qquad i\not= j,\qquad i, j=1,2,3.
\eea

\medskip\noindent
One immediate solution  to (\ref{ee}) consistent with the isotropy of the FLRW metric
is $\w_\mu(t)=(\w_0(t),0,0,0)$. 
There is a second,   ``anisotropic'' solution, with  $\w_{1,2}=0$, $\w_3\not=0$ so
$\w_\mu(t)=(\w_0(t),0,0, \w_3(t))$; then  eq.(\ref{sol}) actually applies to $\w_3$.
This gives a diagonal stress-energy tensor for the contribution of $\w_\mu(t)$
but with  a different value along OZ (as expected).
Since  the contribution of $\w_i$, $i=1,2,3$ to the stress energy tensor in eq.(\ref{eqg})
is suppressed by the scale factor (see (\ref{sol})) and $\w_i$ oscillates rapidly,
the time average of this contribution may be small and  the overall
anisotropy may be  mild enough, while the contribution of $\phi_0$ to the
stress energy tensor may dominate.

Note that,
when taking account of the first equation in (\ref{ww}), then
the first solution above (``isotropic'' case) corresponds
to the limiting case of a Weyl integrable geometry mentioned
earlier, when $\w_\mu$ is ``pure gauge''. The second (``anisotropic'') solution
is the most general in Weyl geometry. The cosmological implications of both solutions are
discussed shortly.

\section{Cosmological applications of Weyl geometry}\label{s3}

The present-day Universe is in a state of accelerating expansion
\cite{1n,2n,3n,4n,Rie,Aad,acc}. The analysis of temperature fluctuations
of the cosmic microwave background radiation
(CMB) by the Planck mission \cite{Pl1,Pl2}
has revealed that the matter content of the Universe consist of $5\%$ baryonic matter and
$95\%$ accounted for by two mysterious components:
the dark energy (with negative pressure) and dark matter
\cite{PeRa03,Pa03,dm1,dm2}, respectively.

To explain these cosmological observations the $\Lambda$CDM model was proposed,
based on the introduction in the Einstein gravitation field equation of the cosmological constant
$\Lambda$, first used by Einstein \cite{Ein} to obtain a static (unstable) model of the Universe.
The $\Lambda$CDM model gives a good fit of the data, but its foundations are questionable due
to the lack of solid theoretical basis; this is due to the uncertainties in the
physical and geometrical interpretation of $\Lambda$ (for a discussion see \cite{Wein,Wein1}).

The Weyl conformal geometry may provide a solution to this problem.
Firstly, we saw in the previous sections that it can naturally recover Einstein gravity and
predicts  a positive cosmological constant in the broken phase.
In this section we examine the implications for cosmology of Weyl quadratic gravity
in its  symmetric phase, together with its underlying Weyl  geometry, and compare the results
to those in  the $\Lambda$CDM model.

Our study below considers first the Weyl model with the  solution that
is compatible with the isotropy of the FLRW metric, i.e. $\w_\mu(t)=(\w_0(t), 0, 0, 0)$.
In the Appendix we re-do the  analysis below for the Weyl model using the second solution 
$\w_\mu(t)=(\w_0(t), 0, 0, \w_3(t))$, and provide the technical details; the formalism
is similar and   in this case we  gain a good insight into the
impact of the effect of space-like components of $\w_\mu$ relative to the isotropic solution,
in a first approximation\footnote{
Strictly speaking, this case would also  demand a suitably modified (``anisotropic'')
FLRW metric along OZ, but
that would introduce an additional parameter (scale factor) in the theory and that would 
make the analysis less predictable.
Including this  case in the analysis here  was motivated by  recent results in  \cite{SS} that
may  question the usual FLRW metric assumption and, secondly,  by  the fact that the
formalism is similar to that of the main, isotropic case.}. The numerical results of the two
cases  will then be compared to those of the $\Lambda$CDM,
see later (Figures~\ref{fig1} and \ref{fig3}).

\subsection{Accelerated expansion}

From eq.(\ref{eqg}) we find from the ``00'' and ``ij'' components, respectively
\bea\label{q1}
\frac{\dot a^2}{a^2}
+\frac{\kappa}{a^2}
-\frac{\ddot \phi_0}{\phi_0}
+3 \,H\, \frac{\dot \phi_0}{\phi_0}
-\frac{\phi_0^2}{12}=0,
\\[3pt]
\frac{\dot a^2}{a^2} + 2\,\frac{\ddot a}{a}
+\frac{\kappa}{a^2}
+
3\frac{\ddot \phi_0}{\phi_0}
+
9 H\, \frac{\dot\phi_0}{\phi_0}
-\frac{\phi_0^2}{4}
=0,\,\label{q2}
\eea
The last term in eqs.(\ref{q1}), (\ref{q2}) is due to the potential
of $\phi_0$.
From eq.(\ref{Keq}) 
\bea\label{4K}
-\frac{\ddot \phi_0}{\phi_0}=\frac{\dot \phi_0^2}{\phi_0^2}
+3 H\,\frac{\dot\phi_0}{\phi_0},
\eea
Then  eqs.(\ref{q1}) and (\ref{q2}) become
\bea\label{q1p}
\frac{\dot a^2}{a^2}
+\frac{\kappa}{a^2}
+
\frac{\dot\phi_0^2}{\phi_0^2}
+
6\,H\, \frac{\dot \phi_0}{\phi_0}
-\frac{\phi_0^2}{12}=0
\\
\frac{\dot a^2}{a^2} + 2\,\frac{\ddot a}{a}
+\frac{\kappa}{a^2}
-
3\frac{\dot \phi_0^2}{\phi_0^2}
-\frac{\phi_0^2}{4}
=0,\label{q2p}
\eea
Subtracting these
\bea\label{sub}
\frac{\ddot a}{a}
-2\frac{\dot \phi_0^2}{\phi_0^2}
-3 H\,\frac{\dot\phi_0}{\phi_0}
-\frac{\phi_0^2}{12}=0.
\eea

\medskip
This shows there is a time-dependent $\ddot a(t)$  of the Universe expansion.
There are three terms contributing to $\ddot a$:
the terms depending on $\dot\phi_0$ are due to $\w_0$; of these,
the term $\propto \dot\phi_0^2$ gives a positive contribution to
$\ddot a$, while the term $\propto\dot\phi_0$ may give a positive (negative) contribution, depending
on the positive (negative) sign of $c_0$ in eq.(\ref{solp}), respectively.
This means it depends on the initial condition
imposed on  $\phi_0(0) \dot\phi_0(0)$.
Further, the  term involving $\phi_0^2$ is due to the potential of $\phi_0$ 
and gives a positive contribution to $\ddot a$ - it is related  to the cosmological constant
$\Lambda=1/4 \langle\phi_0^2\rangle$ after symmetry breaking.

In conclusion, the acceleration is controlled by  $\w_0(t)\sim \partial_0\ln\w_0$
and the scalar mode $\phi_0$ of $\tilde R^2$ term, and is thus of geometric origin.
It is intriguing to see the multiple role of $\phi_0$: it induces the Stueckelberg mechanism
of symmetry breaking and subsequently becomes part of the Weyl-Proca massive field;
its vev generates $M_p=\langle\phi_0^2\rangle/(6\xi^2)$, the cosmological constant
$\Lambda=\langle\phi_0^2\rangle/4$ and
an acceleration of the expansion, giving a dark energy - like contribution.

Let us also consider the limit $t\ra \infty$, then
$\phi_0\ra \langle\phi_0\rangle$  (broken phase),
then from eq.(\ref{q1p})
\bea\label{L1}
\frac{\dot a^2}{a^2}+\frac{\kappa}{a^2}
-\frac13 \, \Lambda\approx 0,
\qquad
\Lambda\equiv\frac14 \langle\phi_0^2\rangle.
\eea
In the same limit, from (\ref{sub})
\bea\label{L2}
\frac{\ddot a}{a}-\frac13\, \Lambda
=0.
\eea
Hence, in this limit the acceleration is  given  by the cosmological constant itself. We simply recovered
from the Weyl model the usual de Sitter exponentially  expanding Universe.

For completeness, let us also present the form of
eqs.(\ref{q1}), (\ref{q2}) in the presence of matter
\bea\label{11}
\frac{\dot a^2}{a^2}
+\frac{\kappa}{a^2}
+\frac{\dot \phi_0^2}{\phi_0^2}
+6 \,H\, \frac{\dot \phi_0}{\phi_0}
-\frac{\phi_0^2}{12}=\frac{1}{3}\frac{6\,\xi^2}{\phi_0^2}\,T_{00}
\\
\label{22}
\frac{\dot a^2}{a^2}
+ 2\,\frac{\ddot a}{a}
+\frac{\kappa}{a^2}
-\frac{\dot \phi_0^2}{\phi_0^2}
+2 \frac{\ddot \phi_0}{\phi_0}
+6 H\, \frac{\dot\phi_0}{\phi_0}
-\frac{\phi_0^2}{4}
=\frac{1}{3}\frac{6\xi^2}{\phi_0^2}\,T_i^i.
\eea
using (\ref{4K}). With $T_{00}$, $T^i_i$ denoting the stress energy tensor matter contributions.
  Eqs.(\ref{11}), (\ref{22}) have similarities to
  eqs.(27), (28) of \cite{AM} which where written without actually providing
  a Lagrangian, by additional assumptions, in
  an attempt to  uplift the Einstein's equations to a scale invariant form
(the coefficient  6 in (\ref{11}) and eqs.(\ref{22}) is replaced in \cite{AM} by 2 and 4
respectively and  $\phi_0\ra \lambda$).

\subsection{  Friedmann equations and  Weyl cosmology}

The general  Friedmann equations are shown in  eqs.(\ref{q1p}), (\ref{q2p}).
Using eq.(\ref{solp}) we replace the derivatives of $\phi_0$ in terms of $\phi_0(t)$ itself.
Then
\medskip
\be\label{Fr3}
3H^{2}=
-\frac{3 \,c_0^{2}}{a^{6}\phi _{0}^{4}}
-18H\frac{c_0}{a^{3}\phi _{0}^{2}}+\frac{\phi _{0}^{2}}{4},
\ee
and
\be\label{Fr4}
2\dot{H}+3H^{2}=
\frac{3\,c_0^{2}}{a^{6}\phi _{0}^{4}}
+\frac{\phi _{0}^{2}}{4}.
\ee

To study the cosmology of the  model defined by eqs.(\ref{Fr3}) and (\ref{Fr4}) 
we first  re-express  the time coordinate, the Hubble function and
  $\phi_0$ in terms  of dimensionless variables $( \tau,h,\phi)$
  \medskip
  \begin{equation}
  \tau =H_{0}\,t,\quad
  H=H_{0}\,h,\quad
  \phi _{0}=H_{0}\,\phi,\quad
\label{4}
\end{equation}%

\medskip\noindent
where $h$, $\phi$ 
are functions of $\tau$; $H_{0}$ is the present value of the Hubble function.
 Therefore
 \medskip
\begin{equation}\label{5}
\dot\phi_0(t)=\phi^\prime( \tau)\,H_0^2,
 \qquad
 \dot\phi_{0}(t)\vert_{t=0} =\phi^{\prime }(\tau)\vert_{\tau=0}\,H_{0}^{2},
 \qquad
 \phi_{0}(t)\vert_{t=0}=\phi(\tau)\vert_{\tau=0}\, H_{0}.
\end{equation}
From eq.(\ref{solp}) 
\bea
\label{Fr5}\qquad\qquad\qquad
\frac{d\phi }{d\tau }-\frac{c_h}{a^{3}\phi}=0,\qquad \textrm{where}\quad
c_h\equiv \frac{c_0}{H_0^3}=\big[\phi(\tau)\phi'(\tau)\big]\vert_{\tau=0}.
 \eea
The Friedmann equations  become
\bea \label{Fr7}
&& 3\,h^{2}\left( \tau \right)=
  -\frac{3\,c_h^2}{a^{6}\phi^{4}}
   -\frac{18\,c_h\, h}{a^{3}\phi^{2}}
+\frac{\phi^{2} }{4},\quad\,\,\, (=\rho_{\textrm{eff}}).
\\[10pt]
\label{Fr6}
 && 2\frac{dh(\tau) }{d\tau }+3h^{2}(\tau)
  =
\frac{3 \,c_h^2}{a^{6}\phi^{4}}+\frac{\phi^2}{4},\qquad (=-p_\eff).
\eea

\medskip
To compare the cosmological predictions of the Weyl geometry-based  model
to the  $\Lambda$CDM and to the observations, we introduce the redshift $z$
via\footnote{As a result, we have
$\label{tz}
{d/dt}=(dz/dt)(d/dz)=-(1+z)\,H(z) \,(d/dz).
$}
\,\,$1+z=1/a.$\,
With this
notation and  eqs.(\ref{Fr5}) to (\ref{Fr6})  we have a system of first order
differential equations of the cosmological evolution in the redshift space
\bea\label{Frf1}
\frac{d\phi(z)}{dz} +(1+z)^{2}\frac{c_h}{\phi(z) \,h(z)}&=&0,
\eea
 with  $c_h=-\phi(z\!=\!0) \phi^\prime(z\!=\!0)$ and
\bea
\label{Frf2}
(1+z)^3\,\frac{d}{dz}\, \Big\{\frac{h^2(z)}{(1+z)^3}\Big\}\,
+\frac{3\, c_h^2\, (1+z)^5}{\phi^{4}(z)}
+\frac{\phi^2(z)}{4 (1+z)}=0,
\eea

\medskip\noindent
and  the constraint (closure relation)
\medskip
\bea\label{Frf3}
\hspace{-0.5cm}
h^{2}(z)
+
\frac{6\, c_h\,(1+z)^3}{\phi^{2}\left(z \right) }
\,  h(z)
+
\frac{c_h^2\,(1+z)^6}{\phi^{4}\left( z\right)} 
-
\frac{\phi^{2}\left( z\right) }{12}=0.
\eea

\medskip\noindent
Eqs.(\ref{Frf1}) to (\ref{Frf3}) define our Weyl cosmological model.
This is  solved numerically for various initial conditions
$(\phi(z=0),\phi^\prime(z=0),  h(z=0))$.
From  eq.(\ref{Frf3}), one can also express analytically
$h(z)$ in terms of $\phi(z)$ (or vice-versa) and replace it in (\ref{Frf2}). At large field values
$\phi(z)^6\gg 96 c_h^3 (1+z)^6$, the middle terms in (\ref{Frf3})  are suppressed and then 
$ h(z)\approx \phi(z) /(2 \sqrt{3})$.

Finally,  introduce the deceleration function $q(z)$
\bea\label{qz}
q\!=\!\frac{d}{d\tau }\frac{1}{h(\tau )}-1
= (1+z)\frac{1}{h(z)}\frac{dh(z)}{dz}-1,
\eea

\medskip\noindent
with  $h(z)$  a solution to the above system.
Eq.(\ref{qz}) will be used  for the numerical analysis.

With the notation in eqs.(\ref{Fr7}), (\ref{Fr6}), we find
\bea
6h(\tau )\frac{dh(\tau )}{d\tau }=\frac{d\rho _{\textrm{eff}}(\tau )}{d\tau }.
\eea
and
\bea
\frac{d\rho _\eff(\tau )}{d\tau }+3h(\tau )\big[\, p_\eff\left( \tau
\right) +\rho _\eff(\tau )\,\big] =0.
\eea
This  gives the energy conservation  equation for the Weyl cosmological model.

\subsection{Weyl cosmology  versus  $\Lambda$CDM}

In this section we compare the $\Lambda$CDM model to the Weyl cosmological model
defined above. In the $\Lambda$CDM model the
simplifying hypothesis that the matter content of the late Universe contains
only dust matter is generally adopted. Therefore the matter in the present day Universe has
negligible thermodynamic pressure. Hence, the energy conservation equation,
$\dot{\rho}+3H\left(\rho+p\right)=0$,
of standard cosmology gives for the time variation of
the energy density of the dust  matter with $p=0$ the simple expression $\rho= \rho
_0/a^3=\rho _0 (1+z)^3$, where $\rho _0$ is the present day matter density.

The time evolution of the Hubble function in terms of the
scale factor and of the redshift $z$ is given  by \cite{e8}
\bea
H\!&=&\! \!H_0\sqrt{\left(\Omega _b+\Omega _{DM}\right)a^{-3}+\Omega _{\Lambda}}
\nonumber\\[6pt]
\!&=&\! \! H_0\sqrt{\left(\Omega _b+\Omega _{DM}\right)(1+z)^{3}+\Omega _{\Lambda}},\qquad
\eea
where  $\Omega _b$, $\Omega _{DM}$, and $\Omega _{\Lambda}$ denote
the density of the baryonic matter, of the cold (pressure-less)
dark matter, and of the dark energy (modeled by a cosmological constant),
respectively, while $H_0$ is the present-day value of the Hubble function\footnote{
  For the Hubble constant we take the value $H_0=67.74\pm 0.46$ km/s/Mpc \cite{Pl1,Pl2}.}.
These three densities obey the closure relation $\Omega _b+\Omega
_{DM}+\Omega _{\Lambda}=1$, which shows that  the geometry of the Universe is
  flat, a relation that was confirmed by observations \cite{Pl1,Pl2}.

The deceleration parameter in the standard  general relativistic cosmology is
given by
\medskip
\begin{equation}
q(z)=\frac{3 (1+z)^3 \left(\Omega _{DM}+\Omega _b\right)}{2 \left[\Omega
_{\Lambda}+(1+z)^3 \left(\Omega _{DM}+\Omega _b\right)\right]}-1.
\end{equation}

\medskip\noindent
To compare the predictions of the  Weyl cosmological model
to the $\Lambda$CDM  model, we adopt for the density parameters the  values
$\Omega_{DM}= 0.2589\pm 0.0057$, $\Omega _{b}= 0.0486\pm 0.0010$, and $\Omega _{\Lambda}=0.6911\pm 0.0062$,
respectively, which follow from the Planck data \cite{Pl1,Pl2}.
Hence, the total matter density $\Omega _m=\Omega _{DM}+ \Omega _b\approx 0.31$.
With the help of the density parameters we obtain
for the present-day value of the deceleration parameter the value $q(0)=-0.5381$.
This   indicates that at present the Universe is in an accelerating phase.
In our comparison below of the  Weyl model versus $\Lambda$CDM
we also  include the observational data for the redshift dependence of
the Hubble function, by using the data quoted in Table IV of \cite{Bou}
(see references therein for the observational results and their error bars).

\begin{figure}
  \centering
  \includegraphics[width=7cm,height=6cm]{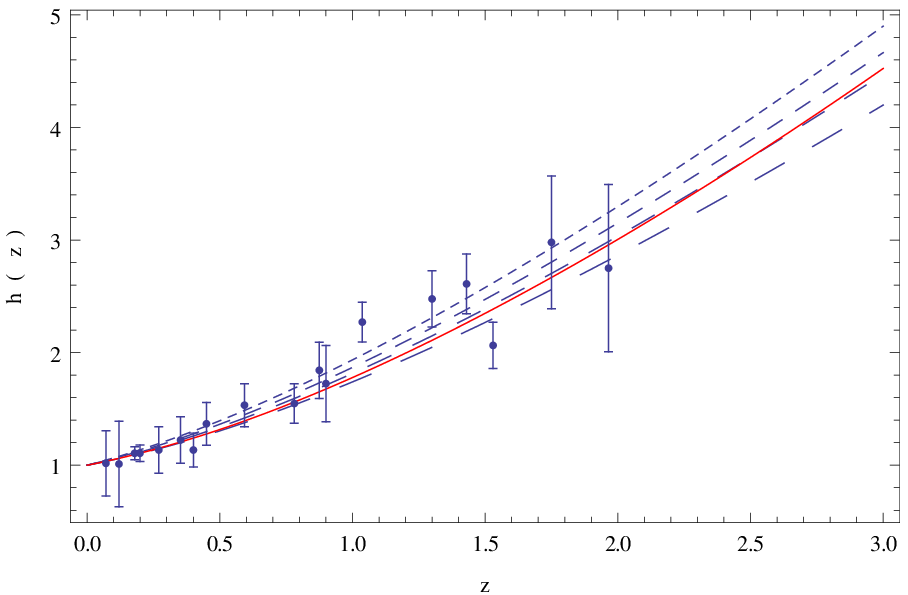}\qquad
   \includegraphics[width=7cm,height=6cm]{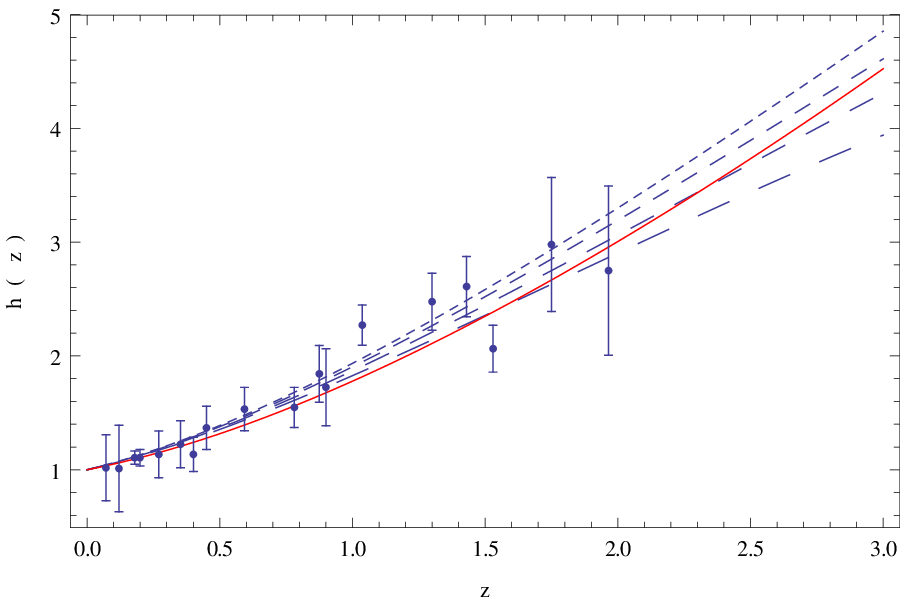}
  \caption{\small
{\bf Left plot}: The case $\w_\mu=(\w_0,0,0,0)$.   The dimensionless
    Hubble function $h(z)$ in $\Lambda$CDM (red curve)
    and in Weyl cosmology as a function of the redshift
    for initial conditions: $h(0)=1$,  $\phi^\prime(z\!=\!0)=0.06$ and  with different
    $\phi(z\!=\!0)=2.67$ (dotted curve),
    $\phi(z\!=\!0)=2.75$ (short dashed curve), $\phi(z\!=\!0)=2.81$ (dashed curve) and $\phi(z\!=\!0)=2.89$
    (long-dashed curve).
    \\
    \noindent
    {\bf Right plot}: The case $\w_\mu=(\w_0,0,0,\w_3)$ (detailed in the Appendix).
    The dimensionless
    Hubble function $h(z)$ in $\Lambda$CDM (red curve)
    and in Weyl cosmology as a function of the redshift
    for initial conditions:  $h(0)=1$,
 $\phi^\prime(z\!=\!0)=0.06$, $\phi(z\!=\!0)=2.81$,
 $\tilde\w(z=0)=4.3$ giving  $\lambda=0.057/\tilde\w(z=0)^2=0.003$,
 for  different values of $u(z=0)$:  $-0.7$ (dotted curve),
 $-0.8$ (short dashed curve), $-0.9$ (dashed curve) and  $-1$
 (long dashed curve). Here $\tilde \w$ and $\w_3$ are related by (\ref{4}).
The experimental data \cite{Bou} are  presented with their error bars.}
\label{fig1}
\bigskip
\includegraphics[width=7cm,height=6cm]{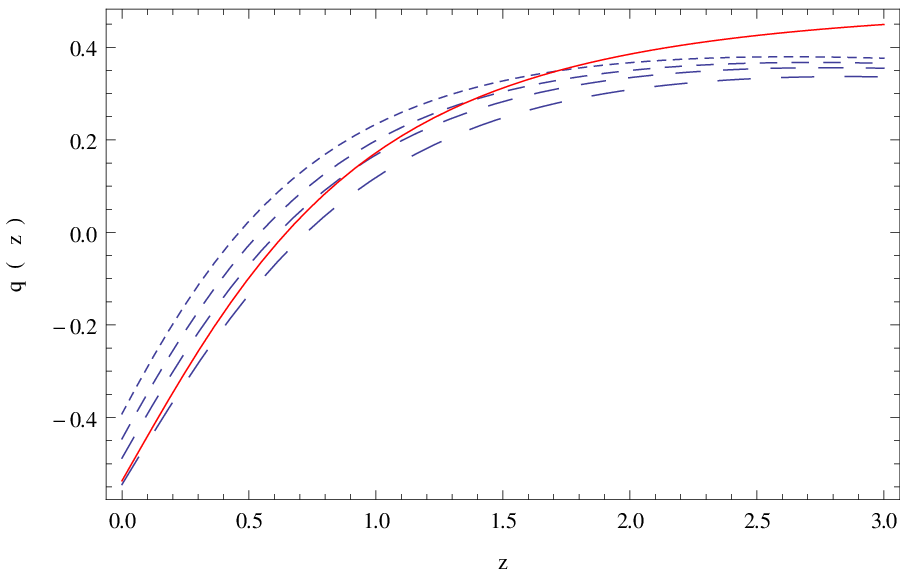}\qquad
\includegraphics[width=7cm,height=6cm]{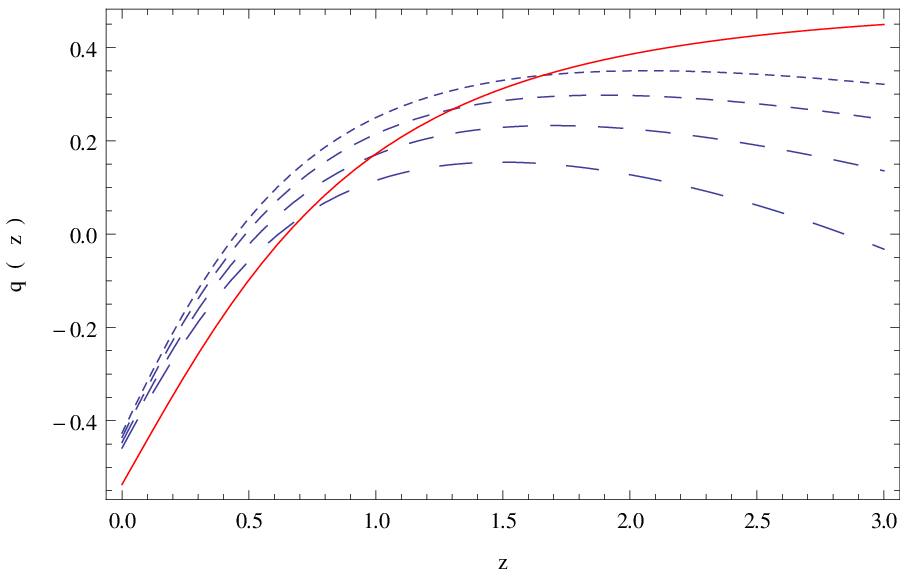}
\caption{\small
 {\bf Left  plot:} The case $\w_\mu=(\w_0,0,0,0)$.
  The  deceleration  $q(z)$ in $\Lambda$CDM (red  curve)
  and in Weyl cosmology  as a function of the redshift $z$ 
  for initial conditions $h(z=0)=1$,  $\phi^\prime(z=0)=0.06$, with $\phi(z=0)=2.67$ (dotted curve),
    $\phi(z=0)=2.75$ (short dashed curve), $\phi(z=0)=2.81$ (dashed curved) and $\phi(z=0)=2.89$
    (long-dashed curve).
    \\
    \noindent
    {\bf Right plot}: The case $\w_\mu=(\w_0,0,0,\w_3)$ (detailed in the Appendix):
    The  deceleration  $q(z)$ in $\Lambda$CDM (red  curve)
    and in Weyl cosmology  as a function of  redshift $z$
    for initial conditions:
    $h(0)=1$, $\phi^\prime(z=0)=0.06$,  $\phi(z=0)=2.81$ and
    $\tilde\w(z=0)=4.3$, hence  $\lambda=0.057/\tilde\w(z=0)^2=0.003$,
 for  different values of $u(z=0)$:  $-0.7$ (dotted curve),
 $-0.8$ (short dashed curve), $-0.9$ (dashed curve) and  $-1$
 (long dashed curve); $\tilde \w$ and $\w_3$ are related by (\ref{4}).
 The deceleration  can return to negative values at higher $z$.
}\label{fig3}
\end{figure}

In Figure~\ref{fig1}, the Hubble function $h(z)$ for the Weyl cosmological
model is compared to its evolution in the $\Lambda$CDM standard
model and to the  data \cite{Bou}.  For a chosen set of initial conditions
shown in this figure,  we see that the Weyl model gives a very good description of the data and
is in  good agreement with  the $\Lambda$CDM model. This is true for the main case
considered here  of solution $\w_\mu=(w_0,0,0,0)$, but also for
the case with solution $\w_\mu=(\w_0,0,0,\w_3)$ discussed in the Appendix,
for suitable initial conditions of the fields detailed in this figure.

In Figure~\ref{fig3} a comparison is shown  of the
deceleration function $q(z)$  in the Weyl cosmological model versus that in the  $\Lambda$CDM.
One can notice that there is a good agreement  between the predicted
evolution of $q(z)$ by the Weyl  model  and the $\Lambda$CDM
for the main case here  with  solution  $\w_\mu=(w_0,0,0,0)$; for the case
with solution $\w_\mu=(\w_0,0,0,\w_3)$ differences emerge for larger values of $z$,
that depend on the initial conditions for the fields; for example,  the deceleration can
return to negative values  near $z\sim 3$.

 The numerical  analysis also shows that the
scalar field $\phi(z)$  is a monotonically increasing function of the redshift
(monotonically decreasing function of  time), in a range between 2.5 and 3.5 (for
the considered $\phi(0)$ values shown in the figures) and for $0\leq z\leq 3$. A similar
(monotonically increasing) behaviour and values
exist for $\tilde \w(z)$ in the case $\w_\mu=(\w_0,0,0,\w_3)$,
with  $0\leq z\leq 3$ and with the corresponding initial conditions of this case
shown in  Figures~\ref{fig1},~\ref{fig3}.

Unlike for the case of $\w_\mu=(\w_0,0,0,0)$, the ``anisotropic'' case $\w_\mu=(\w_0,0,0,\w_3)$
brings in a dependence of the predictions discussed above, on the couplings  $\q$ and $\xi$.
We checked that these couplings remain in  a  perturbative regime for a suitable choice of $\w_3(0)$
and $\w_3^\prime(0)$, see the Appendix.

\section{Conclusions}\label{s4}

The Weyl conformal geometry is relevant in the early Universe where all states are
essentially  massless and effective theories at short distances may become  conformal or Weyl invariant.
Therefore, this geometry provides the natural framework for studying cosmology.
Theories built in Weyl geometry have the unique feature that
both  the action and the underlying geometry (connection) are Weyl (gauge) 
invariant. Theories built in the Riemannian space-time do not have this feature.
This geometry also allows a natural embedding of the Standard Model {\it without}
any additional degrees of freedom beyond those of the SM and Weyl geometry.
Models based on Weyl geometry have successful inflation with predictions similar to those
of the Starobinsky  model.

With this motivation,  we studied the cosmological evolution of  the Weyl conformal geometry
and its associated  Weyl quadratic gravity, in the absence of matter. In previous work we showed
at the level of the Lagrangian  how the Weyl gauge field becomes massive,
by a Stueckelberg mechanism in which  the Weyl field is ``absorbing'' the massless
scalar field $\phi_0$ present in the $\tilde R^2$ term in the action.
This mechanism is of geometric nature i.e. it takes place in the absence of matter,
since both the Weyl vector $\w_\mu$ and the scalar $\phi_0$ have a geometric origin.
In this work we re-examined this mechanism at the level of the equations of motion.
We showed how the symmetry breaking and the gauge fixing condition specific
to massive  Proca fields $\nabla_\mu\w^\mu=0$ are a natural result of the
cosmological evolution. This relation emerges dynamically in a FLRW Universe,
from the Weyl  current conservation, after $\phi_0$  becomes constant (acquires a vev) at
late times.
This shows  the spontaneous breaking of the Weyl gauge symmetry,  in the absence
of matter, to an Einstein-Proca action. After the massive Weyl gauge field decouples,
the Einstein gravity is recovered with a positive cosmological constant.
The  mass of $\w_\mu$  is  near the Planck scale,
unless one tunes $\q\!\ll\! 1$.

We showed that the Weyl conformal geometry alone  provides a natural explanation for the
accelerated expansion of the Universe. The associated, relevant degrees of
freedom are: the scalar mode $\phi_0$  (that ``linearises''  the $\tilde R^2$ term in the action)
and the time component of the  Weyl gauge field $\w_\mu$ that give positive contributions to this
acceleration; further, $\w_0$ also has a negative (dark matter-like) contribution,
while spatial components $\w_k$, $k=1,2,3$, if present,  have negative contributions, too.
The scalar $\phi_0$ also generates the cosmological constant, that gives a dark energy-like
contribution.

We compared  the Weyl cosmological model
to the data and to the  $\Lambda$CDM model based on Einstein gravity with a cosmological constant.
The Weyl  integrable model,  with solution $\w_\mu=(\w_0,0,0,0)$  consistent with the  FLRW metric,
and $\Lambda$CDM model have a similar dependence of  the Hubble
function $h(z)$ and deceleration function $q(z)$ in terms of the redshift variable
($z\leq 3$). In this case, the  agreement with the $\Lambda$CDM and also with the data is
independent  of the actual values of the couplings $\xi$ and $\q$.
We also explored the more general Weyl model having an anisotropic solution
 $\w_\mu=(\w_0,0,0,\w_3)$, to gain an insight into this case. We found in general
 similar results, for perturbative values of the couplings of the theory.
In conclusion,  Weyl  geometry and its
 associated   quadratic gravity can provide an interesting alternative
to the $\Lambda$CDM and to the Einstein gravity;  this means that, ultimately,
the underlying geometry of our Universe may actually  be Weyl conformal geometry.
These results open a new
direction of research in cosmology that deserves further study.

\section*{Appendix}

\def\theequation{A-\arabic{equation}}
\def\thesubsection{A}
\setcounter{equation}{0}
\def\thefigure{A-\arabic{figure}}
\def\thelabel{A}
\label{AA}

\subsection*{A. Cosmological applications:  second solution}

We present here the implications for cosmology of the  second (``anisotropic'') solution
discussed in Section~\ref{GFSB}, $\w_\mu=(\w_0(t), 0, 0, \w_3(t))$, 
that leads to the numerical results presented in Figures~\ref{fig1} and \ref{fig3}.
The analysis is very similar to that in the text for the isotropic solution.

\subsubsection*{A.1 Accelerated expansion}

From eq.(\ref{eqg}) we find the equations for the ``00'' and ``ij'' components
\medskip
\bea\label{aq1}
\frac{\dot a^2}{a^2}
+\frac{\kappa}{a^2}
-\frac{2\,\xi^2}{\phi_0^2}\, T^+_{\w}
-\frac{\ddot \phi_0}{\phi_0}
+3 \,H\, \frac{\dot \phi_0}{\phi_0}
-\frac{\phi_0^2}{12}=0,
\\[3pt]
\frac{\dot a^2}{a^2} + 2\,\frac{\ddot a}{a}
+\frac{\kappa}{a^2}
+\frac{2\xi^2}{\phi_0^2}\,
T_{\w}^-
+
3\frac{\ddot \phi_0}{\phi_0}
+
9 H\, \frac{\dot\phi_0}{\phi_0}
-\frac{\phi_0^2}{4}
=0,\,\label{aq2}
\eea
with
\bea\label{aTw}
T^\pm_{\w}=
\frac{\q^2\phi_0^2}{8\,\xi^2}\,
\frac{\w_k\,\w_k}{a^2}
\pm
\frac{\dot \w_k\dot\w_k}{2 a^2}.
\eea

\medskip\noindent
$T_\w^{\pm}$ is the contribution of space-like components
$\w_k$ ($k=1,2,3$) to the stress-energy tensor,  while the similar
contribution of $\w_0$ is given  by the two terms involving $\ddot\phi_0$, $\dot\phi_0$, via
eq.(\ref{ww}). The sum above over $k$ is actually restricted to $k=3$.
Setting  $T_\w^\pm=0$ one recovers eqs.(\ref{q1}), (\ref{q2}) and subsequent.
Using eq.(\ref{Keq}), (\ref{4K}) then 
\medskip
\bea\label{aq1p}
\frac{\dot a^2}{a^2}
+\frac{\kappa}{a^2}
-\frac{2\,\xi^2}{\phi_0^2}\, T^+_{\w}
+
\frac{\dot\phi_0^2}{\phi_0^2}
+
6\,H\, \frac{\dot \phi_0}{\phi_0}
-\frac{\phi_0^2}{12}=0
\\
\frac{\dot a^2}{a^2} + 2\,\frac{\ddot a}{a}
+\frac{\kappa}{a^2}
+\frac{2\xi^2}{\phi_0^2}\,
T^-_{\w}
-
3\frac{\dot \phi_0^2}{\phi_0^2}
-\frac{\phi_0^2}{4}
=0,\label{aq2p}
\eea
Subtracting them
\bea\label{asub}
\frac{\ddot a}{a}+\frac{\xi^2}{\phi_0^2}\,\frac{\dot \w_k\,\dot\w_k}{ a^2}
-2\frac{\dot \phi_0^2}{\phi_0^2}
-3 H\,\frac{\dot\phi_0}{\phi_0}
-\frac{\phi_0^2}{12}=0.
\eea

\medskip
There is again a time-dependent $\ddot a(t)$  of the Universe expansion
with an interpretation similar to that in the text after eq.(\ref{sub});
however,  now there is an additional contribution from the
term  $\propto\xi^2$ and due to $w_k$ that gives a negative contribution to $\ddot a$;
this term is suppressed by $a^2$ and by the coupling $\xi\ll 1$ of the $\tilde R^2$ term in
the action. In conclusion, the acceleration is controlled by  $\w_0(t)$
and the scalar mode $\phi_0$ of $\tilde R^2$ term, that later becomes the longitudinal
component of massive $\w_\mu$.
Finally, in the limit of $t\ra \infty$, $\phi_0\ra \langle\phi_0\rangle$  (constant) and
neglecting the scale-suppressed  $T_{\w}^+$ (due to $\w_i$)
then eq.(\ref{aq1p}) and (\ref{asub}) recover eqs.(\ref{L1}) and (\ref{L2}) in the text.
In this particular  limit the acceleration is given  by the cosmological constant and
we  recover the usual de Sitter exponentially  expanding Universe.

\subsubsection*{A.2  Friedmann equations and  Weyl cosmology}

The general  Friedmann equations are shown in  eqs.(\ref{aq1p}), (\ref{aq2p}).
Using again solution (\ref{solp}) we replace the derivatives of $\phi_0$ in terms of $\phi_0(t)$ itself.
With  notation (\ref{aTw}),  the Friedmann equations become
\medskip
\be\label{aFr3}
3H^{2}=\frac{6\,\xi^2}{\phi_0^2}\,T_\w^+
-\frac{3 \,c_0^{2}}{a^{6}\phi _{0}^{4}}
-18H\frac{c_0}{a^{3}\phi _{0}^{2}}+\frac{\phi _{0}^{2}}{4},
\ee
and
\be\label{aFr4}
2\dot{H}+3H^{2}=
-\frac{2\,\xi^2}{\phi_0^2}\,T_\w^-
+ \frac{3\,c_0^{2}}{a^{6}\phi _{0}^{4}}
+\frac{\phi _{0}^{2}}{4}.
\ee

\medskip
To estimate the behaviour  of $T_\w^\pm$ in these equations
consider for a moment the particular case when
$\phi_0$ is replaced by $\langle\phi_0\rangle$, then $\w_3$ satisfies
(\ref{ww2})  with solution (\ref{sol}); then
$T_\w^\pm$ is replaced by its vev
\bea
\langle T_\w^\pm\rangle = m_\w^2 \,\frac{\w_k\,\w_k}{2\,a^2}\pm \frac{\dot\w_k\,\dot\w_k}{2\,a^2}.
\eea
With (\ref{sol}), $\langle T_w^\pm\rangle$ are highly oscillatory for $m_\w^2\gg H^2$, hence
one could  replace $\langle T_w^\pm\rangle$ by its time-averaged value
(denoted with a subscript $t$), to  find
\bea
\langle T_\w^+\rangle_t=(A^2+B^2) \frac{m_\w^2}{2\,a^3},\qquad
\langle T_\w^-\rangle_t=0
\eea

\medskip\noindent
This gives an indication of the behaviour of $T_\w^\pm$. We see
that $\w_k$ has vanishing pressure in this approximation and
it  mimics  the dark matter behaviour.
In general, there are  additional corrections to
$\langle T_\w^+\rangle_t\sim 1/a^6$ and $\langle T_\w^-\rangle_t\sim 1/a^3$,
see e.g. \cite{Aluri}.  In our numerical analysis $\phi_0$ is not replaced by its vev
and we use and integrate numerically eq.(\ref{ww}) (instead of eq.(\ref{ww2}) of
solution (\ref{sol})) and compute exactly  the value of $T_\w^\pm$.

As in the text eq.(\ref{4}), to study  the cosmology of the  model defined by (\ref{aFr3}), (\ref{aFr4}) 
we introduce the dimensionless variables $( \tau,h,\phi,\tilde\w)$
  \medskip
  \begin{equation}
  \tau =H_{0}\,t,\quad
  H=H_{0}\,h,\quad
  \phi _{0}=H_{0}\,\phi,\quad
  \omega_3=\frac{1}{\xi}\,H_0\,\tilde\omega
\label{a4}
\end{equation}%

\medskip\noindent
where $h$, $\phi$, $\tilde\omega$ are functions of $\tau$;
 Therefore
 \medskip
\begin{equation}\label{a5}
\dot\phi_0(t)=\phi^\prime( \tau)\,H_0^2,
 \qquad
 \dot\phi_{0}(t)\vert_{t=0} =\phi^{\prime }(\tau)\vert_{\tau=0}\,H_{0}^{2},
 \qquad
 \phi_{0}(t)\vert_{t=0}=\phi(\tau)\vert_{\tau=0}\, H_{0}.
\end{equation}
Then, from eqs.(\ref{solp}), (\ref{ww}) 
\bea
\label{aFr5}\qquad\qquad\qquad
\frac{d\phi }{d\tau }-\frac{c_h}{a^{3}\phi}=0,\qquad \textrm{where}\quad
c_h\equiv \frac{c_0}{H_0^3}=\big[\phi(\tau)\phi'(\tau)\big]\vert_{\tau=0},
 \eea
 \vspace{-0.5cm}
 \bea\label{adw}
 \qquad
 \frac{d^2 \tilde\omega}{d \tau^2}
+ h \frac{d\tilde\omega}{d \tau}
+\lambda 
  \, \phi\,\,\tilde\omega=0,\qquad \textrm{where}\quad \lambda\equiv \frac{\q^2}{4\xi^2}.
\eea
The Friedmann equations  become
\bea \label{aFr7}
 && 3\,h^{2}\left( \tau \right)=\frac{3}{a^{2}}
  \Big\{ \lambda\,\tilde\omega^2 
  +\frac{1}{\phi(\tau)^2}\Big[
  \frac{d\tilde\omega}{d\tau}\Big]^2\Big\}
 -\frac{3\,c_h^2}{a^{6}\phi^{4}}
   -\frac{18\,c_h\, h}{a^{3}\phi^{2}}
+\frac{\phi^{2} }{4},\quad\,\,\, (=\rho_{\textrm{eff}}).
\\[10pt]
\label{aFr6}
 && 2\frac{dh(\tau) }{d\tau }+3h^{2}(\tau)
  =
-\frac{1}{a^{2}}
  \Big\{ \lambda\,\tilde\omega^2  
  -\frac{1}{\phi^2}\Big[\frac{d\tilde\omega}{d\tau}\Big]^2\Big\}
+
\frac{3 \,c_h^2}{a^{6}\phi^{4}}+\frac{\phi^2}{4},\qquad (=-p_\eff).
\eea

\medskip\noindent
Notice now the dependence of these equations on 
$\lambda$ and thus on the couplings $\q$, $\xi$.

To compare the cosmological predictions
to  $\Lambda$CDM we define 
\be u(\tau)=d\tilde\omega/d\tau.
\ee
and also introduce the redshift $z$ via $1+z=1/a$.
Then eqs.(\ref{aFr5}) to (\ref{aFr6})  show  a system of first order
differential equations of the cosmological evolution in the redshift space
\bea\label{aFrf1}
\frac{d\phi(z)}{dz} +(1+z)^{2}\frac{c_h}{\phi(z) \,h(z)}&=&0,
\\[6pt]
\frac{d\tilde\omega(z)}{d z}+\frac{u(z)}{(1+z)\,h(z)}&=&0,
\\[6pt]
\frac{d u(z)}{d z}- \frac{u(z)}{1+z}-\frac{\lambda\,\phi^2(z)\,\tilde\omega(z)}{(1+z) h(z)}&=&0,
\eea
 with  $c_h=-\phi(z\!=\!0) \phi^\prime(z\!=\!0)$ and
\bea
\label{aFrf2}
(1+z)^3\,\frac{d}{dz}\Big\{\frac{h^2(z)}{(1+z)^3}\Big\}
-\frac{\lambda\,\phi^2(z)\,\tilde\omega^2(z)-u^2(z)}{\phi^2(z)} (1+z)
+\frac{3\, c_h^2\, (1+z)^5}{\phi^{4}(z)}
+\frac{\phi^2(z)}{4 (1+z)}=0\quad
\eea

\medskip\noindent
with  the closure relation
\medskip
\bea\label{aFrf3}
h^{2}(z)
-
\frac{(1+z)^2}{\phi^{2}(z)}
\big[\lambda \phi^2(z)\tilde\omega^2(z)
+u^2(z)\big]
+
\frac{c_h^2\,(1+z)^6}{\phi^{4}\left( z\right)} 
+
\frac{6\, c_h\,h(z)}{\phi^{2}\left(z \right) }
\left( 1+z\right) ^{3} 
\!- \!
\frac{\phi^{2}\left( z\right) }{12}\!=\!0.
\quad\eea

\medskip
The set of eqs.(\ref{aFrf1}) to (\ref{aFrf3}) define the Weyl cosmological model for the second solution
for $\w_\mu$
(compare against eqs.(\ref{Frf1}) to (\ref{Frf3}) of ``isotropic'' solution). This set is  solved numerically
with initial conditions $(\phi(z=0),\phi^\prime(z=0), \tilde\omega(z=0), u(z=0), h(z=0))$.
The results of this investigation are presented in Figures~\ref{fig1} and \ref{fig3} in the text (right plots).

Finally, from (\ref{aFr7}) 
\bea
6h(\tau )\frac{dh(\tau )}{d\tau }=\frac{d\rho _{\textrm{eff}}(\tau )}{d\tau }.
\eea

\medskip\noindent
With notation (\ref{aFr7}), (\ref{aFr6}), we find
\bea
\frac{d\rho _\eff(\tau )}{d\tau }+3h(\tau )\big[\, p_\eff\left( \tau
\right) +\rho _\eff(\tau )\,\big] =0.
\eea
This  gives the energy conservation  equation of the model.

\subsubsection*{A.3  Constraints on the couplings}

One may ask what constraints the  couplings $\xi$ and $\q$ must respect to have
the above agreement(s).  To this purpose, one uses the closure relation eq.(\ref{aFrf3})
for $z=0$, with present-day value $h(0)=1$ and finds a constraint
\medskip
\bea\label{t1}
\lambda\,\tilde\w^2(0)\, \phi^2(0)+ u^2\vert_{z=0}
=
\phi^2(0)\Big\{
1+\frac{\phi^{\prime 2}(0)}{\phi^2(0)}-6
\frac{\phi^\prime(0)}{\phi(0)}-\frac{\phi^2(0)}{12}\Big\}.
\eea

\medskip\noindent
The deceleration function $q(z)$ gives another constraint for $z=0$, from (\ref{qz})
\medskip
\bea\label{t2}
\lambda\,\tilde\w^2(0) \,\phi^2(0)- u^2\vert_{z=0}
=2\phi^2(0) \Big[q(0)-\frac12 \Big]+3\phi^{\prime 2}(0)+\frac14 \phi^4(0).
\eea
The last two equations give
\bea
\lambda\,\tilde\w^2(0)&=&
q(0)+2 \frac{\phi^{\prime 2}(0)}{\phi^2(0)}-3 \frac{\phi^\prime(0)}{\phi(0)}+\frac{1}{12}\phi^2(0),
\\
u^2\vert_{z=0}&=&
\phi^2(0)\Big[1-q(0)-\frac{\phi^{\prime 2}(0)}{\phi^2(0)}-3 \frac{\phi^\prime(0)}{\phi(0)}
-\frac16 \phi^2(0)\Big].\label{u2}
\eea
The initial (present day) conditions for $\tilde \w(0)$ and
its derivative  $u\vert_{z=0}$ are determined by the present-day
values of the scalar field $\phi(z=0)$,
of its derivative $\phi^\prime(z=0)$ and of the deceleration $q(z=0)$.
The  initial value of the Weyl vector ($\tilde\w$) is also related to the ratio
$\lambda=\q^2/(4\xi^2)$; a given value of $\tilde\w(0)$
 is fixing  {\it the ratio} of the couplings $(\q, \xi)$   (and vice-versa).

For example, for generic values
 considered in Figures~\ref{fig1} and \ref{fig3} of
 $\phi(z=0)=2.81$, $\phi^\prime(z\!=\!0)=0.06$ (also  $\tilde\w(z\!=\!0)=4.3$),
 which reproduce the $\Lambda$CDM, then $\lambda\,\tilde \w^2(0)\approx 0.056$ and hence
 $\lambda\approx 0.003$; therefore  $\q<\xi$.
 We also find from (\ref{u2}) that $u^2\vert_{z=0}=1.24$.
 These initial conditions  can be re-formulated as constraints for
 $\w_3(0)$, $\q$, $\xi$,   by using definitions (\ref{a4}), (\ref{adw}) for 
 $\tilde\w$ and $\lambda$:
\bea
\q^2\,\w_3^2(0) \approx 0.22\,H_0^2,\qquad 
\xi^2\,\Big[\frac{d \w_3}{dz}\Big]_{z=0}^2\approx 1.24\,H_0^2,
\eea
where $H_0=2.1978\times 10^{-18} s^{-1}$ (corresponding to $H_0=67.8$ Km/s/Mpc).
The couplings  $\q$ and $\xi$ can thus  be in  a  perturbative regime for a suitable choice of
$\w_3(0)$ and $\w_3^\prime(0)$, as mentioned in the text.

\medskip
\bigskip\noindent
{\bf Acknowledgement:}
This work was supported by a grant of the Romanian Ministry
of Education and Research, CNCS-UEFISCDI, project  PN-III-P4-ID-PCE-2020-2255
(PNCDI~III).

\end{document}